# Sub-micron silicon-on-insulator resonator for ultrasound detection

Rami Shnaiderman [1,2,3]*, Georg Wissmeyer[1,2,3], Héctor Estrada [1,2], Daniel Razansky [1,2]

Qutaiba Mustafa [1,2], Andriy Chmyrov [1,2], Vasilis Ntziachristos [1,2]*

[1] Chair of Biological Imaging and TranslaTUM, Technische Universität München, Munich, Germany

[2] Institute of Biological and Medical Imaging, Helmholtz Zentrum München, Neuherberg, Germany

[3] These authors contributed equally: Rami Shnaiderman, Georg Wissmeyer.

* e-mail: rami.shnaiderman@tum.de , v.ntziachristos@tum.de

**Point-like broadband ultrasound detection can significantly increase the resolution of ultrasonography and optoacoustic (photoacoustic) imaging[1,2], yet current ultrasound detectors cannot be miniaturised sufficiently. Piezoelectric transducers lose sensitivity quadratically with size reduction[3], while optical micro-ring resonators[4] and Fabry–Pérot etalons[5] fail to adequately confine light at dimensions smaller than ~50 microns. Micro-machining methods have been used to generate arrays of capacitive[6] and piezoelectric[7] transducers, but at bandwidths of only a few MHz and dimensions not smaller than 70 microns. Here we use the widely available silicon-on-insulator (SOI) platform to develop the world's smallest ultrasound detector with a sub-micron sensing area of 220 $x$ 500 nanometers. The SOI-based optical resonator design can provide per-area sensitivity that is $10^4$-fold higher than for micro-ring resonators and $10^8$-fold higher than for piezoelectric detectors. We also demonstrate ultra-wide bandwidth reaching 230 MHz and conduct the first imaging based on an SOI ultrasound detector. The technology showcased is suitable for manufacturing ultra-dense detector arrays (>125 detectors/mm$^2$), which have the potential to revolutionise ultrasonography and optoacoustic imaging.**

Ultrasound detection based on optical methods has a fundamental advantage over piezoelectric detection because the detectors can be miniaturised without sacrificing sensitivity[3]. One highly miniaturisable approach for ultrasound detection is the use of optical interferometry with a π-shifted Bragg grating etalon embedded in a fibre waveguide[8]. In this configuration, ultrasound waves perturb an optical cavity established between two Bragg gratings, which act as optical mirrors. The ultrasound waves alter the optical path by changing the cavity's length and refractive index[9], allowing the waves to be detected. However, such etalons are unattractive for biomedical imaging because their large sensing length (100-300 microns)[9,10] and narrow bandwidth (10-30 MHz)[9,10] mean they cannot be used as point-like detectors for high-resolution visualisation. In addition, the long Bragg gratings (as long as 2 mm)[9] result in very long etalons with the cavities that are embedded along the optical axis of the waveguide and detect ultrasound waves arriving perpendicularly to the fibre. This geometrical arrangement and dimensions further limit the ability to miniaturise these etalons or use them as forward-looking detectors.

We introduce a novel concept for ultrasound detection that exploits high-throughput fabrication techniques widely used in the semiconductor industry based on the highly scalable SOI platform. Using this technology, we have designed a point-like 220 x 500 nm$^2$ silicon waveguide-etalon detector (SWED) that is at least four orders of magnitude smaller than the smallest available polymer micro-ring detectors[4], and an order of magnitude smaller than the diameter of cells and common capillaries. We show that the SWED concept features unprecedented ultrasound detection characteristics, despite the miniaturisation achieved.

The novel SWED detector (**Fig. 1a, b**) contains a single continuous silicon waveguide divided into four sections: Au layer, spacer, cavity and Bragg grating (**Fig. 1b**). A ~100 nm thick metallic reflective layer (**'Au'; Fig. 1a, b**) was sputtered onto the polished end facet of the waveguide, followed by a spacer section consisting of an ultra-short Bragg grating approximately ~20 µm long (**'Spacer'; Fig. 1a, b**). The reflective layer together with the spacer act as the first optical mirror of the etalon. By employing an ultrathin metallic layer instead of a Bragg grating, we were able to place the optical cavity in close proximity to the end facet of the waveguide, allowing ultrasound detection through its cross-section. Furthermore, the thinness of this mirror minimises attenuation of the ultrasonic waves that propagate from the reflective layer into the optical cavity. A waveguide segment 320 nm long lies adjacent to the spacer, forming the etalon cavity (**'Cavity'; Fig. 1a, b**). The other end of the cavity is terminated by a second optical mirror made of a second Bragg grating 125 µm long (**'Bragg grating'; Fig. 1a, b**). The cavity's length was designed such that the acquired optical round-trip phase shift is π at the resonance wavelength of the etalon. The Bragg gratings were constructed by adding lateral corrugation (**'Corrugation'; Fig. 1b**) to the waveguide with a corrugation depth of *Δw* (**Fig. 1b,** *bottom right*), periodicity of 320 nm and duty cycle of 50%.

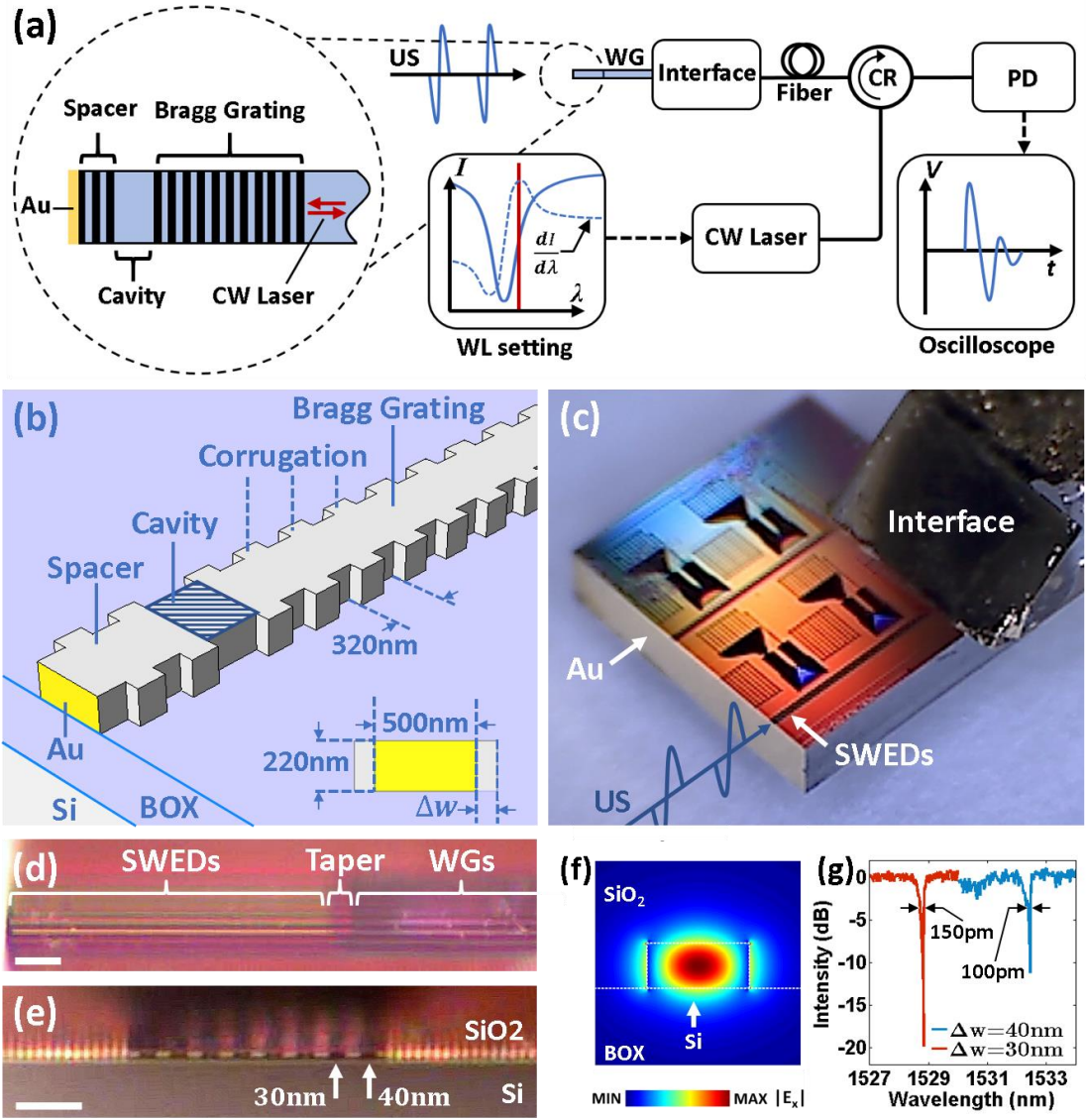

**Fig. 1. Design and operating principle of the SWED. (a)** SWED principle of operation and the back-end systems. CR, fibre circulator; CW, continuous wave; PD, photodiode; WG, waveguide; WL, wavelength; US, ultrasound. The dashed curve in 'WL setting' indicates the slope of the resonance as a function of wavelength, and the red horizontal line indicates the off-resonance tuning of the CW laser. **(b)** Schematic of a single SWED. The corrugation depth on the sides of the Bragg grating is defined as $\Delta w$, depicted at the lower right. BOX indicates the silicon oxide substrate of the silicon waveguide. **(c)** Photograph of the SOI chip with the Au-coated facet facing the ultrasound source. US, ultrasound wave. **(d)** Brightfield micrograph of the SOI chip taken perpendicularly to the optical axis of the SWEDs. WG, waveguide. Scale bar, 20 μm. **(e)** Brightfield micrograph of the SOI chip taken in the direction of the optical axis of the SWED, prior to the application of the Au coating. Scale bar, 20 μm. **(f)** Normalized profile of the horizontal component of the electric field ($E_x$) over an area of 1 x 1 μm. Dashed white lines indicate the boundaries of the waveguide. **(g)** Reflection spectra of SWEDs with corrugation depths of 40 nm and 30 nm, as defined at the lower right of panel (b).

In our implementation, we manufactured eight SWEDs on a single SOI chip measuring 3 *x* 3 *x* 0.8 mm³ (**Fig. 1c**). A brightfield micrograph of the chip taken perpendicularly to the optical axis of the SWEDs (**Fig. 1d**) depicts the eight SWEDs, each connected through a 15-μm adiabatic taper ('**Taper**') to eight Si waveguides ('**WGs**'), leading to an interface ('**Interface**'; **Fig. 1b**, see Methods). The interface is connected on its other side (see **Fig. 1a**) to an array of eight single-mode, polarisation-maintaining fibres ('**Fibre**') connecting each of the SWEDs to the circulator ('**CR**'), laser ('**CW laser**'), and photodiode ('**PD**'). A brightfield micrograph of the chip taken in the direction of the optical axis of the SWEDS (**Fig. 1e**), obtained prior to the application of the Au coating, depicts the cross-sections of the eight SWEDs, which differed in their corrugation depth $\Delta w$. **Fig. 1e** highlights two SWEDs that were the focus of this study, manufactured with $\Delta w$ of 30 (SWED-30) and 40 nm (SWED-40).

The SWEDs are 220 nm thick and 500 nm wide, and they support a well-confined optical transverse electric mode (**Fig. 1f**). The SOI platform offers particularly high index contrast $\Delta n \approx 2.5$ between the cladding and the SWED's cavity materials, whereby common fibre-based etalons offer $\Delta n \approx 0.001$. The high-index contrast enables effective light confinement across the waveguide with optical sub-wavelength cross-sections[11,12], a feature not generally available to implementations with lower $\Delta n$. Due to this high optical mode confinement, detection of ultrasound waves occurs predominantly at the sub-micron-sized SWED cross-section on the Au-coated end facet.

The dimensions of the SWED cavities enable single resonances at near-infrared wavelengths. SWED-40 and SWED-30 resonate at respective wavelengths of 1532.5 nm and 1528.8 nm, with respective resonance FWHM values of 100 pm (Q-factor, $1.5 \times 10^4$) and 150 pm (Q-factor, $1.0 \times 10^4$) (**Fig. 1g**). The higher Q-factor for SWED-40 can be attributed to the higher reflectivity of the Bragg grating due to the deeper corrugation. For ultrasound detection the CW laser pumps light into the SWED cavity. To improve the SWED sensitivity, the laser is tuned off-resonance ('**WL setting**'; **Fig. 1a**) and the polarisation is maintained in the transverse electric orientation by the polarisation-maintaining fibres ('**Fiber**'; **Fig. 1a**). Off-resonance tuning places the pump wavelength at the maximum slope of the resonance curve of the etalon (dashed curve in '**WL setting**'; **Fig. 1a**), ensuring that the optical phase variation in response to incident ultrasound waves ('**US**'; **Fig. 1a**) are amplified by the SWED (see 'Sensitivity' in **Methods**). Therefore, for SWED-40 and SWED-30 the pumping laser was tuned, respectively, to 1532.55 nm and 1528.87 nm. The reflected intensity modulated by the ultrasound waves is detected by a photodiode ('**PD**'; **Fig. 1a**) and recorded by an oscilloscope ('**Oscilloscope**'; **Fig. 1a**).

To characterise the sensitivity and overall operational characteristics of the novel SWED, we employed a calibrated needle hydrophone. An ultrasonic signal was generated via the optoacoustic effect by focusing a pulsed laser (wavelength, 532 nm; pulse width, 0.9 ns) onto a 2.2-µm spot on a piece of black vinyl tape 125 µm thick. The ultrasound source was positioned at a fixed distance first in front the hydrophone and then in front of the SWEDs, using water for acoustic matching (see **Methods**). The noise equivalent pressure (NEP) for SWED-40 and SWED-30 was determined to be, respectively, 60 Pa (12 mPa/Hz$^{1/2}$) and 238 Pa (47.6 mPa/Hz$^{1/2}$) over a 25 MHz bandwidth around the central frequency of 88 MHz (see **Methods**). We observed that a 10-nm increase in corrugation depth translated to ~4-fold improvement in the signal-to-noise ratio (**Fig. 2a**), owing to the increased Q-factor of the cavity.

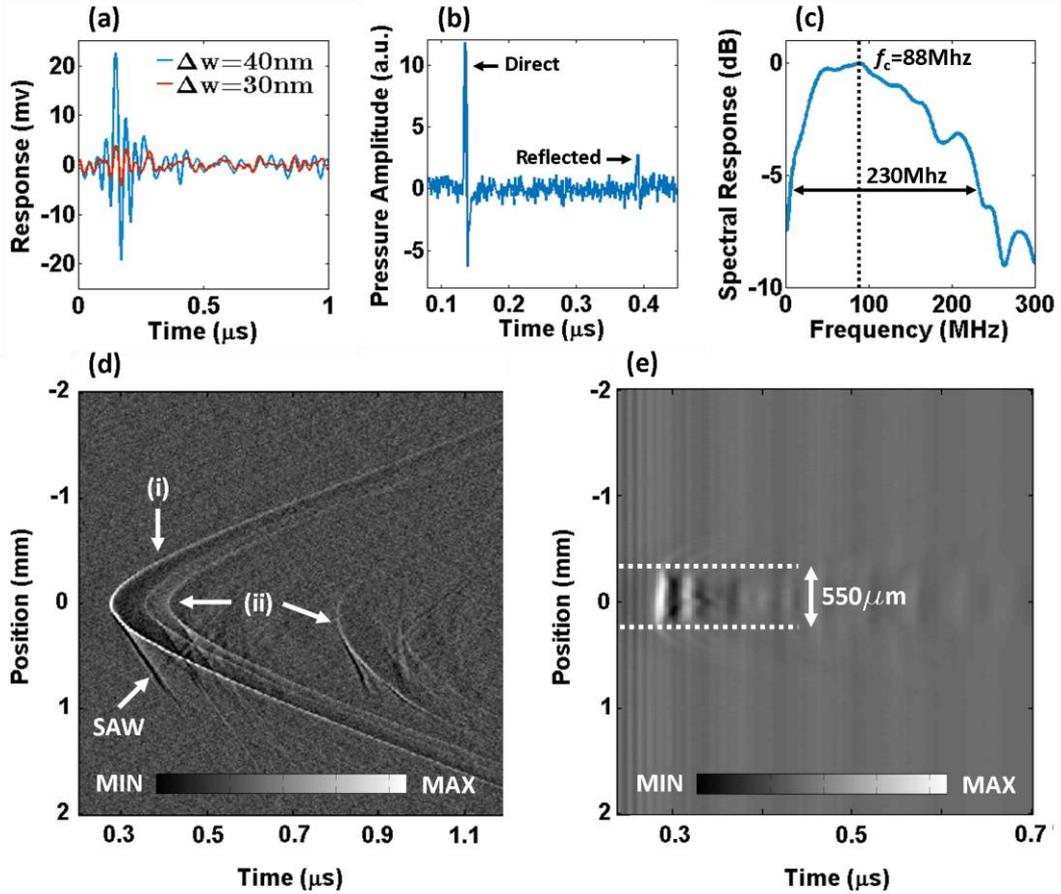

**Fig. 2. Characterization of SWED responses. (a)** Temporal response of two of the SWEDs following exposure to a broadband ultrasound source. The detectors differ in corrugation depth $\Delta w$ **(see Fig. 1b, *lower right*)**. **(b)** Temporal response of the most sensitive SWED-40 showing a signal arriving directly at the detector from the source ("Direct") and another signal arriving to the detector after reflection from the sample holder ("Reflected"). **(c)** Spectral response of SWED-40 to the direct signal (arrow "Direct") in panel (b). **(d)** Spatial response of SWED-40 acquired by scanning the SWED-40 linearly along a distance of 4 mm over a point ultrasound source. Pressure amplitude is depicted on a continuous grayscale (see bottom of image). Arrow "(i)" indicates longitudinal waves, arrow "SAW" indicates surface acoustic waves, and arrows "(ii)" indicate reflections from the sample holder. **(e)** Spatial response of a needle hydrophone with a diameter of 0.5 mm acquired by scanning the hydrophone linearly along a distance of 4 mm over a point ultrasound source. Pressure amplitude is depicted on a continuous grayscale (see bottom of image).

Next, we characterised the SWED-40 detection bandwidth. The SWED was exposed to a broadband ultrasound point source generated by focusing a pulsed laser (wavelength, 532 nm; pulse width, 0.9 ns) onto a 2.2-µm spot on an Au film that had been sputtered to a thickness of 200 nm onto a glass substrate. The film was acoustically matched to the SWED with water. The raw response of the SWED (**Fig. 2b,** arrow "**Direct**") shows a characteristic 'N-shaped' acoustic signal as theoretically expected for a broadband detector. A second recorded signal (arrow "**Reflected**") is a reflection of ultrasound from the glass substrate of the point source. The spectral content of the direct signal (**Fig. 2c**) indicates an ultrasonic detection bandwidth of 230 MHz at -6 dB and a central detection frequency of 88 MHz.

To characterise the SWED-40 spatial response, we employed a linear translation stage to scan the detector along a 4 mm distance (10 µm step size) over the same broadband point source employed in bandwidth characterisations. The optoacoustic signals were recorded as a function of time for each translation step. The spatial SWED response (**Fig. 2d**) exhibits a prominent curved profile (profile "**i**") as expected from a point detector. A linear fit of the time-distance plot of profile "**i**" revealed a respective sound velocity of ~1526 m/s. This velocity is in good agreement with the value expected for propagation of longitudinal acoustic waves in water, indicating that the trajectories correspond to ultrasound waves directly detected by the SWED. A second response, annotated as profile "**ii**" in **Fig. 2d,** is attributed to acoustic reflections between the SWED and the sample holder. Finally, a signal with a dominant dip appeared on the image, arriving at much faster velocities, possibly due to the presence of surface acoustic waves (arrow **"SAW"**). Those waves may form when the longitudinal waves incident on the silicon-water interface around the SWED give rise to shear and Rayleigh waves[13]. Surface acoustic waves propagate at the silicon-water interface at least 4-fold faster than longitudinal waves[10], resulting in an increased average velocity of the detected signals.

The acceptance angle of SWED-40 for longitudinal ultrasonic waves was calculated from profile "**i**" (**Fig. 2d**) by determining the maximum angle whereby signals with intensity of at least -6 dB of the most intense signal in **Fig. 2d** could be detected. SWED-40 was able to detect ultrasound waves over an angle of 148°, corresponding to an acoustic numerical aperture (NA) of 0.96. This value is close to the NA = 1 of a theoretical point detector, due to the much smaller sensing area of a SWED compared to the wavelength of the detected ultrasound waves. The high NA indicates that the use of SWED in imaging applications could lead to implementations with higher lateral resolution than the ones achieved with conventional finite-sized detectors[14]. For comparison purposes we also scanned a needle hydrophone of 0.5 mm diameter over an ultrasound point source. The spatial response of the hydrophone (**Fig. 2e**) shows that common finite-size detectors are dominated by their aperture and the achievable spatial sampling accuracy, and that the acceptance angle is much smaller than for a point detector.

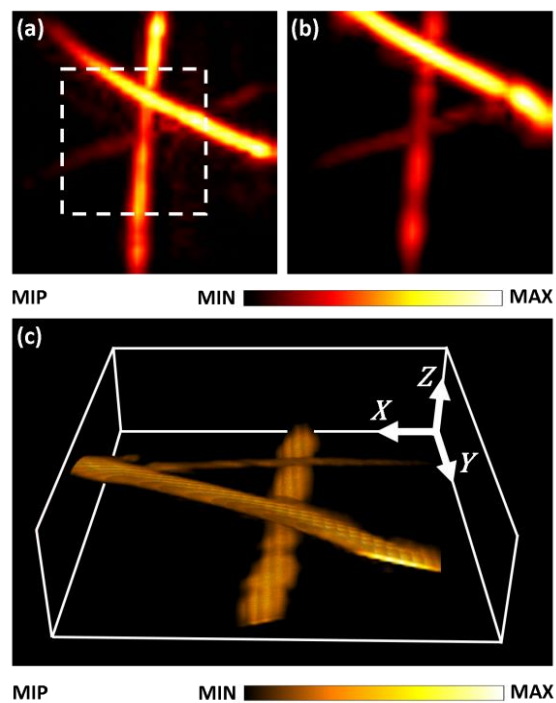

**Fig. 3. Reflection-mode optoacoustic tomography with a SWED** ($\Delta w$ = 40 nm). SWED-40 was used to image a phantom constructed from three black polystyrene sutures with diameters of 10 μm (middle), 30 μm (bottom) and 50 μm (top). The SOI chip was oriented at 45° with respect to the phantom. **(a)** Reconstructed maximum intensity projection (MIP) of a raster scan of the phantom with a step size of 100 μm, covering an area of 4 x 4 mm². **(b)** Reconstructed MIP of the area enclosed by the dashed white line in panel (a), scanned with a step size of 50 μm, covering an area of 2 x 2 mm². **(c)** Volumetric rendering of the data in panel (b) over a depth of 1 mm.

As a first step towards examining whether the SWED-40 can be used for imaging applications, we performed optoacoustic tomography of a phantom constructed from three black polystyrene sutures with diameters of 10, 30, and 50 µm, arranged in a triangular shape. We illuminated this phantom with a pulsed laser (wavelength, 532 nm; pulse width, 0.9 ns) and an optical fluence of 2 mJ/cm². Two raster scans at different scanning steps were performed with the sample at a distance of 1.3 mm from the SWED. The first scan was performed with a step size of 100 µm covering an area of 4 x 4 mm². The three sutures were clearly resolved (**Fig. 3a**) and the highest intensity was observed for the 50-µm suture lying on top of the others. The second scan was performed with a step size of 50 µm over an area of 2 *x* 2 mm², for zoomed-in operation. A reconstructed MIP (**Fig. 3b**) and volumetric rendering of the data (**Fig. 4b**) taken from the area indicated by the dashed white line on **Fig. 4a** showcased the SWED imaging ability for two- and three-dimensional tomography.

## DISCUSSION

We developed the world's smallest ultrasound detector with a sensing area at least 450 times smaller than that of a π-shifted Bragg grating etalon[10] and nearly 2.6 x 10⁴ smaller than that of polymer micro-rings[4]. The new detector yielded highly attractive sensitivity and bandwidth characteristics and was employed to perform the first SOI based imaging ever reported. With dimensions that are 77 and 34 times smaller than the acoustic wavelength at the central detection frequency of 88 MHz, our SWED satisfies the definition of a true point detector.

The product of the NEP and the sensing area allows an objective comparison of detectors with different sizes and operating principles and has been reported to be **1.58×10⁻²** mPa×mm²/Hz^(1/2) for polymer micro-rings employed in optoacoustic microscopy[15] and **509** mPa×mm²/Hz^(1/2) for a miniaturised piezoelectric transducer for intravascular ultrasound imaging[16]. Those values are, respectively, four and eight orders of magnitude lower than the sensitivity achieved by the SWED. This marked sensitivity improvement can be attributed to the high spatial confinement of light in the SOI platform. SWED sensitivity can be further improved by increasing the reflectivities of the Bragg grating and the metallic coating, as well as by reducing side-wall roughness[17] and using rib waveguide geometries[18] to minimise optical losses. Moreover, shortening the spacer and positioning the cavity closer to the waveguide facet may increase both sensitivity and bandwidth.

Combining many SWED elements into one array is technically feasible and could offer new sensing and imaging abilities leading to larger fields of view and faster imaging times than single element detectors. State-of-the-art piezoelectric arrays feature densities of 0.18 detectors/mm$^2$ [19], while arrays of capacitive micromachined ultrasound transducers have been reported with densities of 2.5 detectors/mm$^2$ [20]. Using an SOI array of 8 SWED's, we preliminary managed to increase detection density by orders of magnitude, pointing to the possibility to achieve 125 detectors/mm$^2$. In fact, the technology showcased here could achieve densities of >1000 detectors/mm$^2$ since inter-SWED distances and Si wafer thickness can be further reduced with currently available manufacturing processes.

The high bandwidth and the sub-micron point-like aperture achieved by our SWED offers unique detection performance that may substantially improve axial and lateral resolution in ultrasound and optoacoustic tomography. Increases in bandwidth are essential for achieving high tomographic resolution [14], whereby the sub-micron aperture can lead to unprecedented spatial resolution, setting a new standard for non-invasive ultrasound and optoacoustic imaging [21,22]. The SOI platform may allow the combination of SWED detectors and detector arrays with SOI-based integrated bio-sensors [23] or on-chip microscopes [24], which could lead to powerful integrated tools for basic research and diagnostics.

## METHODS

**Materials and device fabrication**

The chip layout was designed using Optodesigner software (PhoeniX Software; Enschede, Netherlands). It consists of several components: waveguides, grating-couplers, Bragg gratings, and arrayed waveguide gratings. (The arrayed gratings were not used in the present study and are not discussed here.) The chip was fabricated at Interuniversity Microelectronics Centre (IMEC; Löwen, Belgium) through the ePIXfab Consortium Service on an SOI wafer with silicon orientation of (100). The main fabrication techniques included UV-lithography on a standard I-Line resist followed by a two-step etch process of the silicon, involving a shallow etch (70 nm) and a deep etch (220 nm). The components were embedded between 2 µm of $SiO_2$ back oxide (**Fig. 1c**, "BOX") and 1.25 µm of $SiO_2$ cladding. The waveguides were designed to be single-mode with a cross-section of 220 nm in height and 500 nm in width, dimensions commonly used in the silicon photonics field[11]. The Bragg gratings were manufactured by adding lateral corrugations on the waveguides along a length of 250 µm. Adding a discontinuity in the corrugation at the centre of the corrugated section transformed the Bragg gratings into π-shifted Bragg grating etalons.

After manufacture, the wafer was diced into chips measuring 6 *x* 3 *x* 0.8 mm, where the π-shifted Bragg grating etalons are located parallel and centred along the longer dimension of the chip at a distance of 250 µm (**Fig. 1b**, "SWEDs"). The ready-made chip was cut perpendicular to the orientation of the π-shifted Bragg grating etalons at a distance of approximately 500 µm from the centre of the cavity. The chip facet along the cut was then precision-polished with progressively finer diamond-grit polishing films (grit size from 0.1 to 30 µm), followed by a final polish with $SiO_2$ lapping film (grit size 0.02 µm). The spectral responses of the π-shifted Bragg grating etalons were monitored during the polishing process. The polishing was complete once the typical resonance in the spectral response vanished but the bandgap maintained some degree of asymmetry (see the next subsection **Spectral response measurements during the polishing process**). This indicated that one of the sides of the Bragg gratings was nearly entirely polished away, and that the cavity was close to the polished facet. Since the accuracy of the polishing equipment is limited, a buffer zone of ~20 µm was left between the cavity and the polished facet (**Fig. 1c**, "Spacer"). After polishing, the chip measured 3 *x* 3 *x* 0.8 mm and the polished facet was coated with Ti (5 nm) followed by a layer of Au (100 nm) using sputter deposition.

The SWEDs are connected to rectangular Si waveguides measuring 220 *x* 450 nm (**Fig. 1d**, "WGs") via an adiabatically tapered 15 µm section of the waveguides (**Fig. 1d**, "Taper"). These waveguides terminate at an interface that consists of eight grating-couplers, which in turn are coupled to an array of eight single-mode polarisation-maintaining fibres via an in-line coupling element (PLC Connections; Columbus, USA) designed to excite the transverse electric mode.

The facet of the in-line coupling element was also coated with Ti (5 nm) and Au (100 nm) using sputter deposition to ensure efficient coupling from the fibres to the fibre-couplers in water. The fibres in the array act as input/output ports for the interrogation of each SWED and were individually connected to an interrogation scheme based on a tuneable CW laser in the C-band (see **Fig. M2a, b**). The connectorization of the chip was performed by PLC Connections using a fibre array with a pitch of 127 µm.

**Spectral response measurements during the chip polishing process**

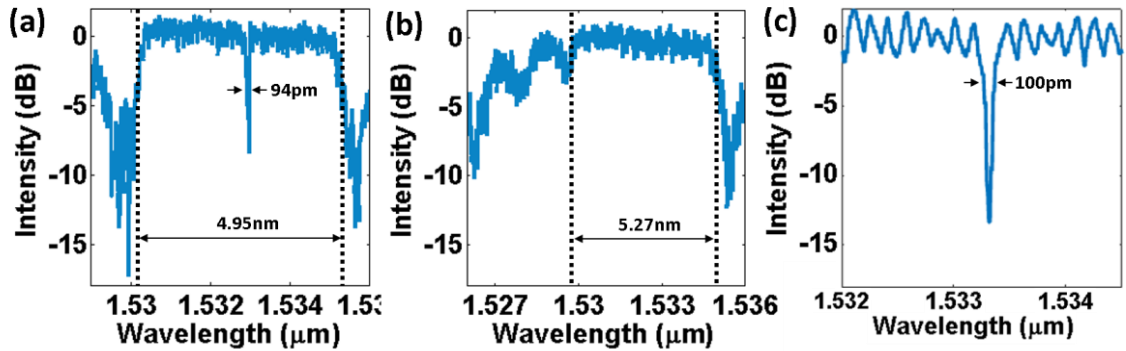

**Fig. M1. Monitoring the reflection spectrum of π-shifted Bragg grating etalon with Δ$w$ = 40 nm during the polishing process of the SOI chip**. **(a)** Before the start of polishing. **(b)** At the end of polishing. **(c)** After application of the reflective coating.

The polishing process has to be carefully monitored in order not to polish away the cavity and at the same time position the cavity as close as possible to the polished facet. It is possible to correlate the shape of the reflection spectra to the length of the Bragg grating polished away from one side of the cavity. **Fig. M1a** shows the reflection spectra of a π-shifted Bragg etalon with Δ$w$ = 40 nm. A characteristic bandgap with a width of ~5 nm (at FWHM) is visible and can be attributed to the Bragg gratings. In addition, there is a resonance 94 pm wide (at FWHM) in the middle of the bandgap attributed to the cavity section. When one of the Bragg gratings is almost entirely polished away, the confinement efficiency of the light in the cavity is drastically reduced and the resonance vanishes, as depicted in **Fig. M1b**. When the facet of the chip is coated with a thin reflective film (5 nm Ti / 100 nm Au), the confinement efficiency improves and a resonance with a similar Q-factor to the original resonance appears (**Fig. M1c**).

**Device characterisation**

During SWED characterization and optoacoustic tomography imaging, ultrasonic signals were detected by monitoring variations in the reflected intensity from the SWED. This was done using a CW-laser (C-band, 20 mW; INTUN TL1550-B, Thorlabs; Newton, USA) and a high-bandwidth photodiode (detection bandwidth, 1.6 GHz; PDB480C, Thorlabs).

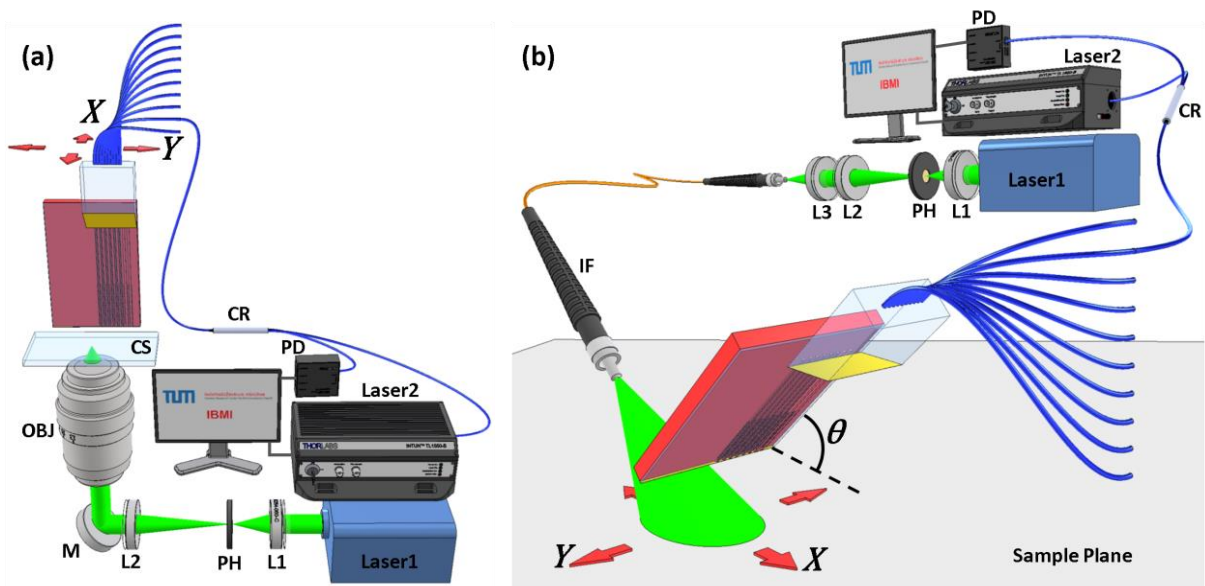

**Fig. M2. Schematic depiction of the experimental setups. (a)** Characterization setup: An inverted microscope with a laser source for optoacoustic optical imaging is combined with the SOI chip mounted in a trans-illumination geometry. The SOI chip is mounted perpendicular to the sample plane and is raster-scanned over it. **(b)** Reflection-mode optoacoustic tomography setup: The laser source for optoacoustic excitation is coupled into an illumination fibre, which is scanned across the sample plane simultaneously with the SOI chip. The angle of the SOI chip relative to the sample plane ($\theta$) can be adjusted. In both setups, the SWED interrogation is performed by a tunable CW laser that can be connected to individual SWEDs. CR, fibre circulator; CS, coverslip; IF, illumination fibre; Laser1, pulsed excitation laser; Laser2, tunable CW interrogation laser; L (1-3), lenses; M, mirror; OBJ, microscope objective; PH, pinhole; PD photodiode.

Ultrasonic signals were excited using a 532-nm pulsed laser with a maximum pulse repetition rate of 2 kHz, pulse width of 0.9 ns and pulse energy of 1 mJ (Wedge HB-532, Bright Solutions SRL; Pavia, Italy). The laser power was attenuated on-demand with neutral density filters inserted along the optical path. A photodiode (DET36A, Thorlabs) located near the laser output triggered signal acquisition by a high-speed 1 GS data acquisition card (GaGe; Lockport, USA).

SWED characterisation was performed using the setup depicted in **Fig. M2a**. The laser beam was guided through a 10-µm pinhole for spatial filtering, resized in a telescope and then focused with a 10x microscope objective (PLN 10x, NA 0.25; Olympus, Germany) with an optical focus of approximately 2.2 µm. The SWEDs were characterised with this ultrasonic point source and acoustically coupled with a few drops of water. The point source was generated when the focal point was aligned to coincide with an optical absorber.

The sensitivity was determined using vinyl black tape 125 µm thick (Type 764, 3M; Neuss, Germany) as optical absorber. The repetition rate was 1 kHz and the pulse energy was 24 µJ. The tape was glued to a microscope coverslip of 150 µm thickness and a 0.5-mm needle hydrophone (Precision Acoustics; Dorchester, UK) was used for calibration of the acoustic source. The spectral response of the needle is approximately constant in the frequency band of [5, 30] MHz and is equal to 450 mV/MPa. SWED response to the tape was recorded without averaging and passed through a [5, 30] MHz band pass filter to spectrally match it with the needle hydrophone for NEP estimation.

For characterisation of bandwidth, a thin gold film (200 nm) was sputtered onto a microscope coverslip in order to serve as optical absorber and ultra-broad bandwidth acoustic source [25]. The chip was positioned on top of the ultrasonic point source using 3D linear translation stages (MTS50-Z8, Thorlabs; not shown in **Fig. M2a**). The repetition rate was 1 kHz and pulse energy was 4.8 nJ. SWED response to the point source was passed through a bandpass filter of [2, 499] MHz and averaged $1 \times 10^4$ times in order to compensate for the extremely weak signals generated from the point source.

The spatial response of the SWEDs was characterised with the same ultrasonic source. The signals recorded at each position were averaged 300 times, and a bandpass filter of [5,150] MHz was applied.

**Noise equivalent pressure**

The needle hydrophone employed for the source calibration in this paper has an approximately constant spectral response in the frequency band of [5, 30] MHz. The NEP of the SWEDs around the central frequency of 88 MHz can be estimated using the following relation:

$$NEP_{min} \cdot R_{max} = NEP_{[5,30]MHz} \cdot R_{[5,30]MHz} \quad (M1)$$

where $NEP_{min}$ and the $R_{max}$ are, respectively, the NEP and spectral response of the SWEDs at 88 MHz; while $NEP_{[5, 30]MHz}$ and $R_{[5, 30]MHz}$ are, respectively, the NEP and spectral response of the SWEDs at the [5, 30] MHz band. The ratio of the spectral responses was calculated from **Fig. 2c** to be ~2.2. The $NEP_{[5, 30]MHz}$ was calculated using the calibrated needle hydrophone.

**Sensitivity**

The sensitivity ($S$) of an optical resonator can be defined as the reflected optical power ($P_R$) modulation per unit of incident acoustic pressure ($p$) [26]:

$$S = \frac{dP_R}{dp} = \frac{dP_R}{d\phi} \cdot \frac{d\phi}{dp} \quad (M2)$$

where $\phi = 4\pi n_{eff} l/\lambda$ is the optical phase, $n_{eff}$ is the effective refractive index of the waveguide, $l$ is the resonator length, and $\lambda$ is the wavelength. The $dP_R/d\phi$ term is the slope of the resonance of the reflected optical power. This term amplifies the detected signal and can be ignored when comparing resonators with identical Q-factors. The term $d\phi/dp$, on the other hand, also called the acoustic phase sensitivity ($S_\phi$), accounts for the properties of the materials from which the detector is constructed as well as for the positioning of the detector relative to the acoustic source. The phase sensitivity can be written as follows:

$$S_\phi = \frac{2\pi n_{eff} l}{\lambda} \cdot \left( \frac{1}{n_{eff}} \frac{dn_{eff}}{dp} + \frac{d\varepsilon_z}{dp} \right) = \frac{2\pi n_{eff} l}{\lambda} \cdot S_\lambda \qquad (M3)$$

where $S_\lambda$ is the normalised sensitivity and $\varepsilon_z$ is the strain along the optical axis. The change in phase can be attributed to two mechanisms: change in $n_{eff}$ of the waveguide due to density changes through the elasto-optic effect, and change in $l$ due to strain experienced by the waveguide.

In order to determine whether ultrasound detection through the cross-section of the SWED (**'forward detection'**) is more efficient than detection perpendicular to the waveguide (**'side detection'**), we compared the normalised sensitivity of the two detection geometries. In the forward detection scheme, the acoustic wave is propagating along the optical axis, so the applied stress is $\sigma_z = P$, and the elastic deformation perpendicular to the optical axis is $\varepsilon_x = \varepsilon_y = 0$. Using the generalised Hooke's law and accounting for the elasto-optic effect in an isotropic cubic material, we obtain the following relationships:

$$\varepsilon_z = -\frac{(1+\nu)(1-2\nu)}{(1-\nu)E} P \qquad (M4)$$

$$\Delta n_x = \Delta n_y = \frac{(\nu C_1 + C_2)}{1-\nu} P \qquad (M5)$$

where $\Delta n_x$ and $\Delta n_y$ are the perturbations of the refractive index perpendicular to the optical axis; $C_1$ and $C_2$ are the elasto-optic constants; E is Young's modulus, and $\nu$ is the Poisson ratio. The values of these parameters for silicon and silica were taken from the literature[27]. The contribution of the strain component to the normalised sensitivity was evaluated directly from **Eq. (M4),** and the contribution due to change in $n_{eff}$ was simulated using **Eq. (M5)** and a vectorial mode solver[28]. For the SWED cross-section in the present study, we found the normalised sensitivity to be $S_{\lambda,forward} = -5.5 \times 10^{-6}$ MPa$^{-1}$. The negative sign implies that the elastic waveguide deformation has a larger impact than the elasto-optic effect. In the side detection scheme, the acoustic wave impedes on the waveguide perpendicularly; hence, $\sigma_y = P$ and $\varepsilon_x = \varepsilon_z = 0$. The normalised sensitivity for the side detection geometry of a waveguide with the same geometrical dimensions and material composition as the SWED was previously found to be $S_{\lambda,side} = 4.7 \times 10^{-6}$ MPa$^{-1}$ [27]. In that case, the positive sign indicates the dominance of the elasto-optic effect.

Comparing the absolute value of the normalised sensitivity of the two detection geometries shows that the forward detection geometry enabled by the SWED is ~20% more efficient than any other SOI-based resonator with similar Q-factor and identical waveguide cross-section dimensions. This normalised sensitivity can be further improved by fine-tuning the dimensions of the cross-section and excitation of other modes[27].

**Surface acoustic waves**

Surface acoustic waves can interfere with image reconstruction, resulting in artefacts due to non-injective mapping of the object. In order to minimise generation of these waves, several spatial orientations of the SOI chip were investigated. **Fig. M3** shows line scans of SWED-40 over an acoustic source after averaging 300 times and applying a bandpass filter of [5, 100] MHz. The chip was tested at various tilt angles (defined in **Fig. M2b**), and scans were performed along the *x* axis (**Fig. M3a, c, e**) and *y* axis (**Fig. M3b, d, f**). We examined the shape of the spatial response and the average propagation velocities of the slowest trajectories indicated by the dashed line. When the tilt angle was 0° or 90°, spatial responses were asymmetrical and dominated by surface acoustic waves with average velocities of 1780 m/s **(Fig. M3a)**, 2890 m/s **(Fig. M3b)**, 2750 m/s **(Fig. M3e)** and 3320 m/s **(Fig. M3f)**. In contrast, a tilt angle of 45° led to a symmetrical spatial response with average velocities of 1860 m/s (**Fig. M3c**) and 1850 m/s (**Fig. M3d**). The results for the tilt angle of 45° indicate that the ultrasound waves propagated for most of the time as longitudinal waves in water. Consequently, the component of the surface acoustic waves contributing to the average velocity is negligible.

These results suggest that surface acoustic waves detected by the SWED can be reduced by minimising the solid angle of the Si-water interface around the SWED, as viewed from the source. Analysis of the raw data in **Fig. M3c** without averaging or bandpass filtering shows that at tilt angle of 45° the SWED has a detection angle of 124° corresponding to an NA=0.88.

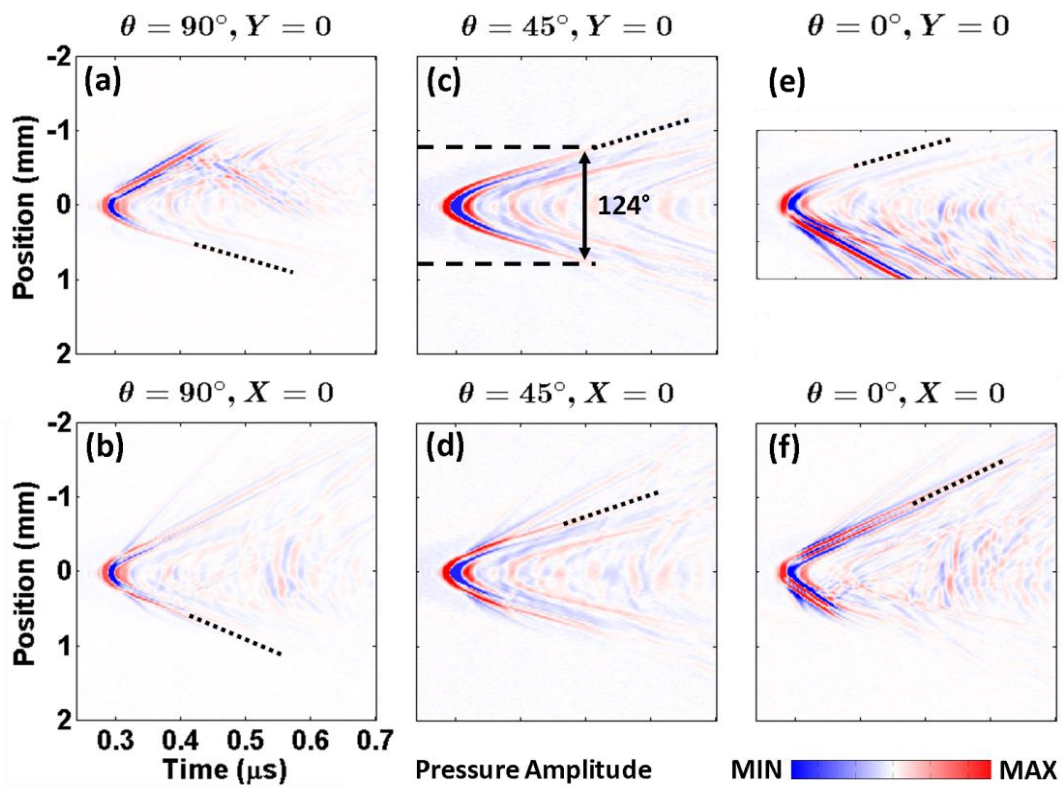

**Fig. M3. Spatial response of the SWED-40.** In each case, the chip was oriented at a tilt angle $\theta$ relative to the sample plane (see **Fig. M2b**) and an ultrasonic source was located at (0,0). The SOI chip was oriented such that the longest dimension of the facet exposed to ultrasound was parallel to the *y* axis (see **Fig. M2b**). **(a, c, e)** Line scans 4 mm along the *x* axis at the indicated tilt angles. In panel (e), it was not possible to scan along more than 2 mm because of physical limitations of the setup. **(b, d, f)** Line scans 4 mm along the *y* axis at the indicated tilt angles.

**Optoacoustic tomography**

For reflection-mode optoacoustic tomography, the beam was guided through a 50-µm pinhole for spatial filtering and then collimated and coupled into a low NA multimode fibre (NA 0.22, core diameter 200 µm; Thorlabs). The illumination fibre as well as the chip were mounted on a set of linear *xyz* translation stages (MTS50-Z8, Thorlabs; not shown in **Fig. M2b**) and simultaneously raster-scanned over the phantom. The chip was positioned at a 45° angle relative to the sample plane to suppress surface wave generation. The sample plane was illuminated by the beam diverging from the fibre, forming a spot of around 2 mm in diameter below the chip. The repetition rate was 2 kHz and the pulse energy was 60 µJ, corresponding to an optical fluence of 2 mJ/cm² which was well below the ANSI limit of 20 mJ/cm² as the maximum permissible visible-light laser fluence for human skin [29].

The phantom was constructed by collocating black polystyrene sutures with diameters of 10, 20 and 50 µm (Dafilon Polyamide, B. Braun Melsungen AG; Melsungen, Germany) in the shape of a triangle. In order to avoid immediate acoustic reflections, the resulting structure was bonded onto a microscope coverslip at a height of 150 µm with UV-curable adhesive (Norland, Thorlabs). The phantom was acoustically coupled to the SWED with water. The acquired raw data were averaged 50 times for enhanced contrast and filtered using a bandpass filter of [1, 35] MHz. The image was then reconstructed using a back-projection method in Fourier space[30].

## **Acknowledgements**

The authors thank A. Chapin Rodríguez, PhD for helpful comments and discussions on the manuscript. The authors would like to thank Josef Promoli (Helmholtz Zentrum München, Germany) as well as the staff at IMEC (Löwen, Belgium), ePIXfab (Ghent, Belgium), PLC

## Author Contributions

R.S. conceived the SWED principle of operation and designed the chip layout. G.W. conceived the design and construction of the SWED in the SOI platform. R.S. and G.W. conducted the experiments. R.S. conducted the data processing and analysis. G.W. designed and constructed the optical setups. R.S wrote the control and acquisition code. Q.M. contributed the image reconstruction code. H.E. and D.R. contributed to surface wave discussion and analysis of the data. A.C. advised on the experimental setup. R.S., G.W. and V.N. wrote the manuscript. V.N. supervised the research. All authors read and edited the manuscript.

## Author Information

Reprints and permissions information is available at www.nature.com/reprints. The authors have no conflicts of interest to declare.